\documentclass[prb,preprint]{revtex4-1} 

\usepackage{amsmath}  
\usepackage{amsfonts} 
\usepackage{graphicx} 
\usepackage[usenames]{color}

\begin{document}


\title{Estimating the settling velocity of fine sediment particles at high concentrations}

\author{Agust\'in Millares}
\affiliation{Andalusian Institute for Earth System Research, University of Granada, Avda. del Mediterr\'aneo s/n, Edificio CEAMA, 18006, Granada, Spain} 
\author{Andrea Lira-Loarca}
\author{Antonio Mo\~nino}
\author{Manuel D\'iez-Minguito}
\email{mdiezm@ugr.es}


\date{\today}

\begin{abstract}
This manuscript proposes an analytical model and a simple experimental methodology to estimate the effective settling velocity of very fine sediments at high concentrations. A system of two coupled ordinary differential equations for the time evolution of the bed height and the suspended sediment concentration is derived from the sediment mass balance equation. The solution of the system depends on the settling velocity, which can be then estimated by confronting the solution with the observations from laboratory experiments. The experiments involved measurements of the bed height in time using low-tech equipment often available in general physics labs. This methodology is apt for introductory courses of fluids and soil physics and illustrates sediment transport dynamics common in many environmental systems.
\end{abstract}

\maketitle 

\section{Introduction} 
The estimation of the terminal fall velocity (settling velocity) of particles within a fluid under gravity is a subject of interest among a wide range of scientific disciplines in physics, chemistry, biology, and engineering. The terminal fall velocity calculation from Stokes' Law is usually taught in introductory courses of physics of fluids as an approximate relation, at low Reynolds numbers, for the drag force exerted on an individual spherical body moving relative to a viscous fluid \citep{mawhinney2012measuring, manson1977stokes}.  

\begin{figure}
\centering{{\includegraphics[scale=0.40]{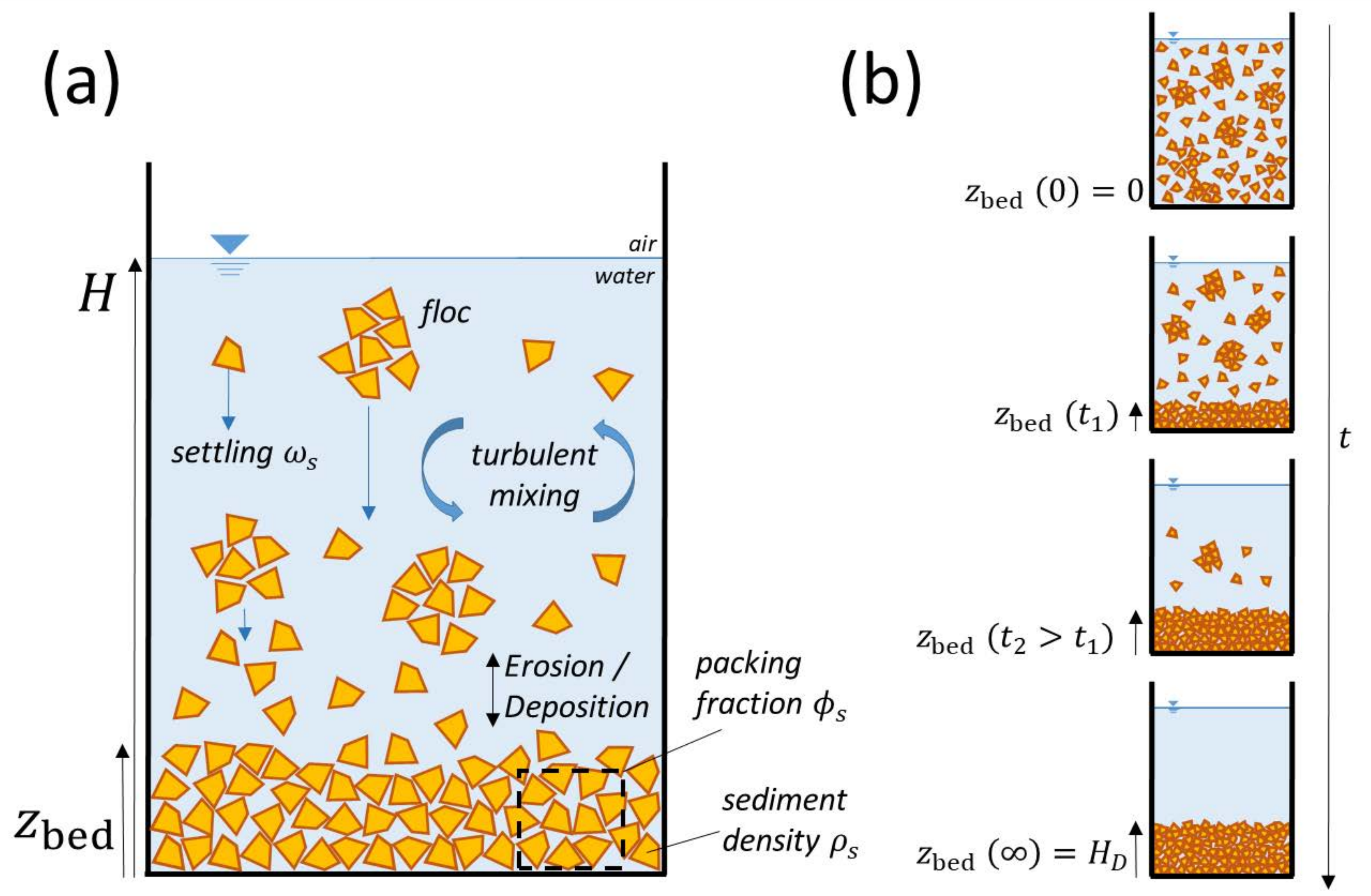}}}
\caption{\footnotesize{(a) Processes and variables definition. The blue down arrows illustrate the different magnitudes of settling velocity. (b) Bed height evolution from an initial state of suspended sediment ($t=0$) up to a fully-deposited sediment bed ($t=\infty$).}} \label{fig:esquema0}
\end{figure}

However, the calculation of the terminal velocity of fine sediment particles (e.g., sand or mud) at high concentrations presents additional issues (Fig.~\ref{fig:esquema0}, panel (a)). Besides their irregular shape, which conditions the drag force and the settling velocity, particles become cohesive and the electrostatic interactions among them and with the ambient fluid allow for flocs formation \citep{winterwerp2004introduction}. Flocs, which are formed by aggregation of sediment particles, typically exhibit larger settling velocities than the primary particles. Moreover, neighboring particles hinder the settling velocity of individual sediment grains, decreasing the overall effective settling velocity of the suspension. Hindered settling, which occurs at concentrations of a few $\mathrm{mg/cm^3}$, is caused by the wakes left by falling particles or flocs, collisions, and increase of effective viscosity, among other processes \citep{winterwerp2004introduction,yin2007hindered}. These issues prevent the use of the Stokes' Law and complicate the basic sediment laboratory experiments usually done in the first years of different bachelor's degrees. Accurate estimates of the effective settling velocity of sand and mud at high concentrations are usually inferred from vertical concentration profiles measured with particle tracking or particle image velocimetry (PTV/PIV) techniques and X-ray devices.  

Therefore, to avoid these issues and with a view to introduce the students to basic sediment transport processes, an analytical model is derived and an experimental methodology is proposed to estimate the effective settling velocity of suspended sediments. 

\section{Analytical Model}
The derivation of an equation for the settling velocity for fine sediment particles starts with the mass balance equation for suspended sediments, which can be expressed as
\begin{eqnarray}
\label{eq:balance}
\frac{\partial c}{\partial t}+ \text{\textbf{div}}\,\mathbf{F}=0\,,
\end{eqnarray}
where $\text{\textbf{div}}\,\mathbf{F}=\sum_{n=\{x,y,z\}}\frac{\partial F_n}{\partial n}$. This equation states the rate of variation of the suspended sediment concentration, $c$ (here in $\mathrm{g\,cm^{-3}}$), with the three-dimensional divergence of sediment fluxes, $F_{n}$ ($\mathrm{g\,cm^{-2}\ s^{-1}}$), in a given control volume. The sediment fluxes $F_n$ through the control volume walls, which comprise advective and diffusive contributions, are 

\begin{eqnarray}\label{eq:massbalancefluxes0}
    F_x = u\,c-K_h\frac{\partial c}{\partial x}\,,\\ \nonumber
    F_y = v\,c-K_h\frac{\partial c}{\partial y}\,,\\ \nonumber
    F_z = (w-\omega_s)\,c-K_v\frac{\partial c}{\partial z}\,, \nonumber
\end{eqnarray}
with $\{x,\,y,\,z\}$ the three Cartesian directions, and $(u,\,v,\,w)$ the components of the three-dimensional current vector. Each sediment flux is comprised by the advection due to the fluid current and gravity and the diffusive contribution due to turbulent mixing, which is commonly treated as a Fickian process. The coefficients $K_h$ and $K_v$ ($\mathrm{cm^{2}\ s^{-1}}$) are, respectively, the horizontal and vertical turbulent diffusion coefficients, and $\omega_s$ is the effective settling velocity ($\mathrm{cm\ s^{-1}}$).

Assuming that the ambient fluid is at rest and that the movement of the sediment particles occurs mostly due to gravity along the vertical coordinate $z$ (i.e., $F_z\gg F_x\,,F_y$ or horizontally homogeneous control volume), Equation~\ref{eq:balance} reduces to 
\begin{eqnarray}
\label{eq:eqvertical}
\frac{\partial c}{\partial t}\approx \frac{\partial F_z}{\partial z} = \frac{\partial}{\partial z}\left(\omega_s c+K_v\frac{\partial c}{\partial z}\right)
\,,
\end{eqnarray}
where the advective flux term, $\omega_s c$, only depends on the effective sediment settling velocity, assumed constant for simplicity. In particular, $z=0$ is set at the base of the control volume, positive upwards (Figure~\ref{fig:esquema0}). The integration of Equation~\ref{eq:eqvertical} in a water column of height $H$ (Fig.~\ref{fig:esquema0}, panel (a)) requires one initial condition and two boundary conditions. The boundary conditions read
\begin{eqnarray}
\label{eq:BC1}
F_z(H)=-\omega_s c(H)-K_v\frac{\partial c(H)}{\partial z}=0\,,\\ 
\label{eq:BC2}
F_z(z_{\text{bed}})=-\omega_s c(z_{\text{bed}})-K_v\frac{\partial c(z_{\text{bed}})}{\partial z}=E-D\,,
\end{eqnarray}
where $z_{\text{bed}}$ is the bed height and $c(z_{\text{bed}})$ is the sediment concentration just above bed. Equation~\ref{eq:BC1} states that a zero- sediment flux condition is applied at the water free surface. The flux condition at the bed is determined by the balance between sediment erosion and deposition rates, namely, $E$ and $D$, respectively (Eq.~\ref{eq:BC2}).

Equation~\ref{eq:eqvertical} is now integrated from $z=z_{\text{bed}}$ to $z=H$. Then, imposing the boundary conditions (Eqs.~\ref{eq:BC1}-\ref{eq:BC2}), the depth-integrated equation reads
\begin{eqnarray}
\label{eq:eqverticalintegrada}
\frac{d \left(H-z_{\text{bed}}\right)\overline{c}}{d t}=E-D\,,
\end{eqnarray}
with $\overline{c}(t)=1/\left(H-z_{\text{bed}}\right)\int_{z_{\text{bed}}}^{H}c(z,t)\,dz$ the time-dependent depth-averaged suspended sediment concentration. The balance between erosion and deposition rates, $E-D$, determines the net sediment concentration within the water column. Erosion and deposition rates depend on the bed shear stress induced by currents and the critical shear stresses for erosion and deposition (e.g., \cite{einstein1962experiments,mehta1988laboratory}). If the ambient fluid is assumed at rest, the bed shear stress is negligible and the erosion rate at the bed is zero, thus only sedimentation occurs. The deposition rate $D$ is then estimated as $D\approx \omega_s c(z_{\text{bed}})$ \citep{sanford1993assessing}. Therefore, Equation~\ref{eq:eqverticalintegrada} reduces to 
\begin{eqnarray}
\label{eq:eqverticalintegrada2}
\frac{d \left(H-z_{\text{bed}}\right)\overline{c}}{d t}=-\omega_s \overline{c}\,,
\end{eqnarray}
assuming for simplicity that $c(z_{\text{bed}})\approx \overline{c}$, which is a reasonable approximation when the water column is considered well-mixed. 

As the sediment concentration in the water column decreases, the bed height increases (Fig.~\ref{fig:esquema0}, panel (b)). The bed height variation rate, which also depends on the deposition rate $D$ and the sediment volume fraction (i.e., packing fraction) in water $\phi_s$, reads 
\begin{eqnarray}
\label{eq:dzdt}
\frac{d z_{\text{bed}}}{d t}=\frac{D}{\rho_s \phi_s}=\frac{\omega_s \overline{c} }{\rho_s \phi_s}\,,
\end{eqnarray}
where $\rho_s$ is the sediment density (e.g., $\rho_s\approx 2.65\,\mathrm{g/cm^3}$ for quartz particles). Since only sedimentation is assumed, $dz_{\text{bed}}/dt$ is positive. Notice also that $z_{\text{bed}}$ increases at a faster rate when $\omega_s$ and $\overline{c}$ present larger values. The lower the sedimentation rate, the slower $z_{\text{bed}}$ increases. When the sediment packing fraction at the bed is more efficient (larger $\phi_s$) the bed height varies more slowly. 

Equations~\ref{eq:eqverticalintegrada2} and \ref{eq:dzdt} represent a (non-linear) system of two coupled Ordinary Differential Equations (ODEs) for $\overline{c}(t)$ and $z_{\text{bed}}(t)$, respectively. An experimentally convenient initial condition is that all the sediment is in suspension, i.e., $z_{\text{bed}}(0)=0$ and $\overline{c}(0)=M/\left(A\,H\right)$, where $M$ is the total mass of sediment and $A$ is the horizontal cross-section of the container ($\mathrm{cm^2}$). When all the sediment is deposited, the bed height is $z_{\text{bed}}(\infty)\equiv H_D$ (Fig.~\ref{fig:esquema0}, panel (b)). In that situation, 
\begin{eqnarray}
\label{eq:HD}
M=\rho_s\phi_s H_D A\,.
\end{eqnarray}
This equation can be used to estimate the volume fraction of the sediment $\phi_s$ from $M$, $H_D$, and $A$ which can be easily measured. 

\begin{figure}
\centering{{\includegraphics[width=12cm]{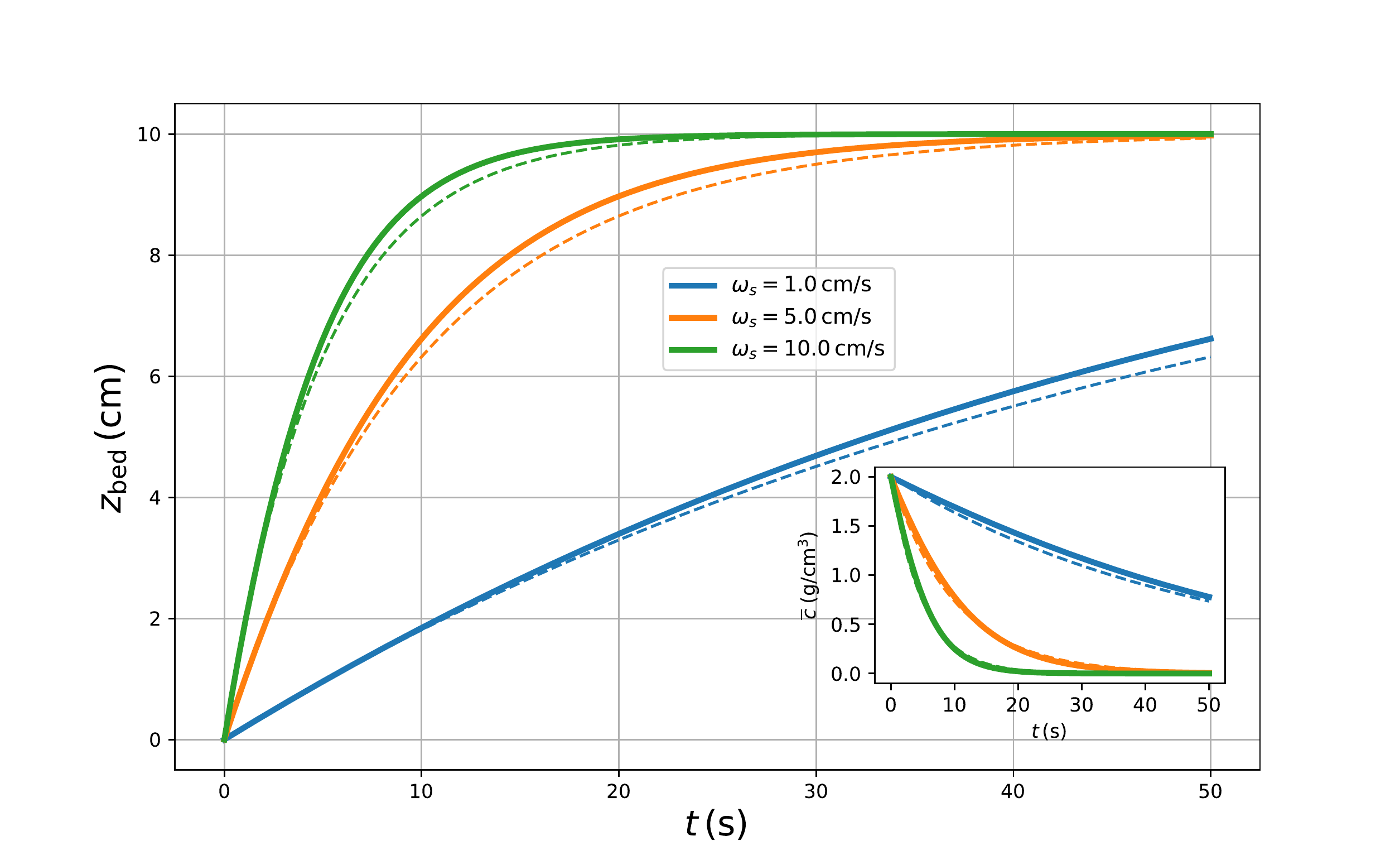}}}
\caption{\footnotesize{Numerical solution of the non-linear system of ODEs~\ref{eq:eqverticalintegrada2} and \ref{eq:dzdt} (thick solid lines) for $\omega_s=1\,\mathrm{cm/s}$, $\omega_s=5\,\mathrm{cm/s}$, and $\omega_s=10\,\mathrm{cm/s}$. The main graph shows the bed height evolution and the inset the depth-averaged suspended sediment concentration. For the three cases considered, $H=50\,\mathrm{cm}$, $H_D=10\,\mathrm{cm}$ and $c_D=M/A/H_D=10\,\mathrm{g/cm^3}$. The thin dashed lines are the corresponding approximate solutions from Eqs.~\ref{eq:approxC} and \ref{eq:approx}, i.e., assuming $H_D\ll H$.}} \label{fig:example_sol}
\end{figure}


Regarding the non-linear system of ODEs~\ref{eq:eqverticalintegrada2} and \ref{eq:dzdt}, its integration provides the sought time evolution for $z_{\text{bed}}(t)$. A numerical solution is determined here from the Python code '\textit{settling.py}' provided within the Supplementary Material, which is described in Appendix A. Figure~\ref{fig:example_sol} shows the solutions for the three different settling velocities indicated in the legend. As expected, the results showed in the main graph indicate that the bed height increases faster (slower) when $\omega_s$ is larger (smaller). Conversely, the suspended sediment concentration decays at lower rates when the settling velocity value is smaller (inset). Approximate solutions of the system are also shown in Figure~\ref{fig:example_sol} (thin lines). These simpler analytical solutions can be derived considering that $z_{\text{bed}} \ll H$. This approximation linearizes Equation~\ref{eq:eqverticalintegrada2}, whose solution is therefore
\begin{eqnarray}
\label{eq:approxC}
\overline{c}=\overline{c}(0)\exp\left(-t/\tau\right)\,,
\end{eqnarray}
with $\tau=H/\omega_s$ the $e$-folding time or time scale for the sediment concentration decay in the water column. The larger the $e$-folding time, the larger the time the sediment takes to be deposited (inset in Fig.~\ref{fig:example_sol}). For the settling velocities of  $\omega_s=1\,\mathrm{cm/s},\,5\,\mathrm{cm/s}$, and $10\,\mathrm{cm/s}$, the $e$-folding time is $50\,\mathrm{s},\,10\,\mathrm{s}$, and $5\,\mathrm{s}$, respectively (Fig.~\ref{fig:example_sol}). 

Finally, Eq.~\ref{eq:dzdt} can be readily integrated by using Eq.~\ref{eq:approxC} to obtain an approximation for bed height evolution. This reads  
\begin{eqnarray}
\label{eq:approx}
z_{\text{bed}}=H_D\left(1-\exp\left(-t/\tau\right)\right)\,.
\end{eqnarray}

\begin{figure}
\centering{{\includegraphics[width=13cm]{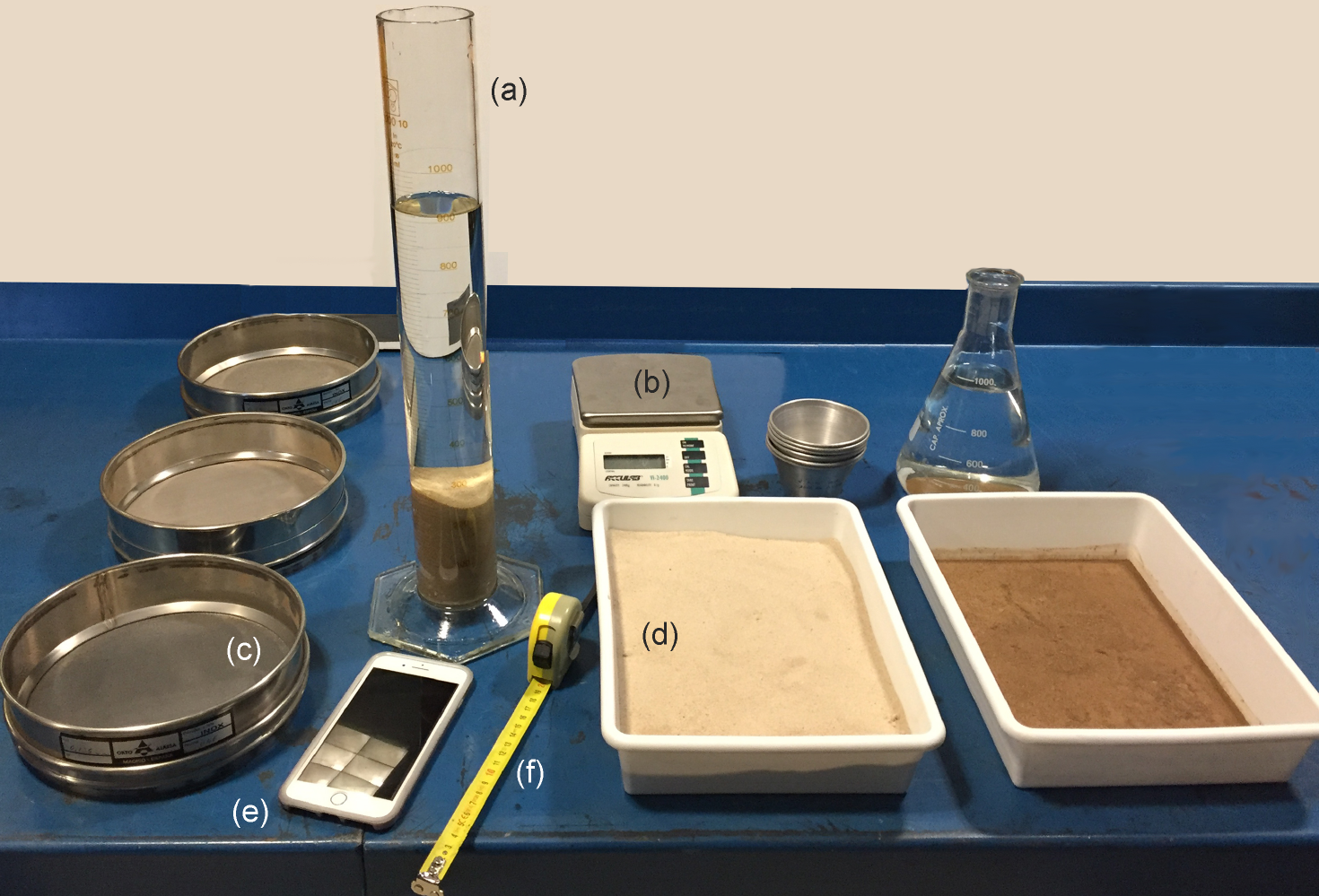}}}
\caption{\footnotesize{Overall picture of the lab materials needed for the experiment: (a) Graduated cylinder with water, (b) Weight scale, (c) Sieves of different mesh sizes, (d) Sediment sample(s), (e) Chronometer and video options from a smartphone, (f) Measuring tape.}} \label{fig:materials}
\end{figure}

\begin{figure}
\centering{{\includegraphics[width=7cm]{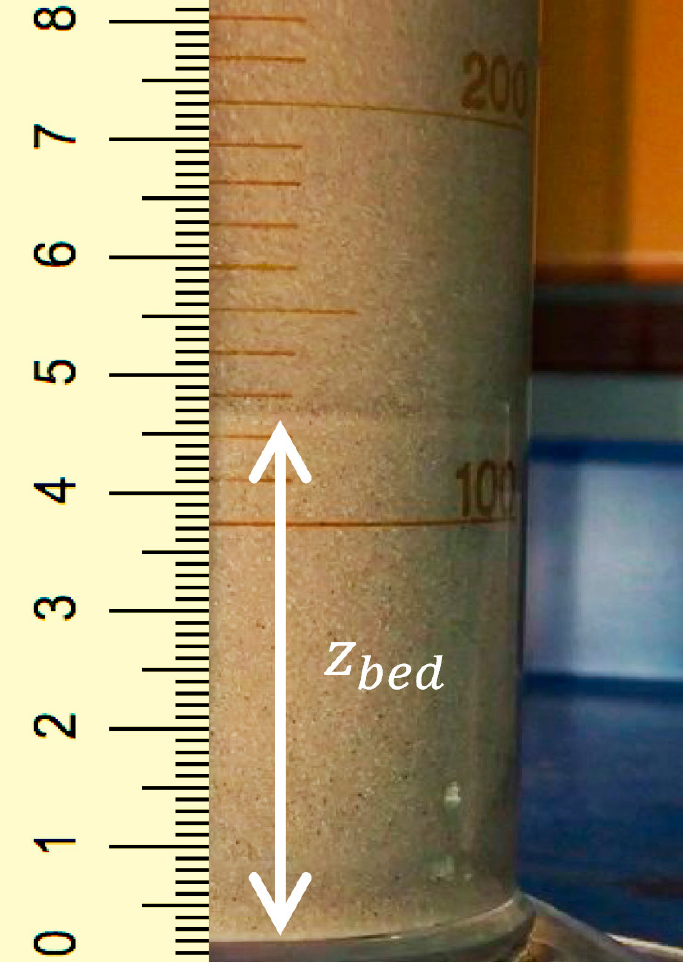}}}
\caption{\footnotesize{Picture of suspended sediments and bed height at a certain time during one of the experiments.}} \label{fig:experiment}
\end{figure}

\section{Experimental Procedure and Results}\label{sec:procedure}

The settling velocity can be estimated by confronting the exact solution for $z_{\text{bed}}(t)$ with the observations of the bed height as a function of time measured in the lab. Based on the analytical model solutions, the following procedure is proposed to carry out experiments to estimate the settling velocity $\omega_s$ (and also the packing fraction $\phi_s$):
\begin{enumerate}

\item[1.] Prepare the following materials (Figure~\ref{fig:materials}).

	\begin{itemize}
	\item Graduated cylinder or other transparent rigid container with uniform cross-section.
	\item Weight scale.
	\item Measuring tape.
	\item Chronometer.
	\item Sample of well-sorted fine sediment (e.g., fine sand or silt will work). Different samples of different sizes would also allow for comparisons. 
	\item Water. 
	\item Sieves of different mesh sizes (optional).	
	\end{itemize}
\item[2.] Measure the cross-section $A$ of the container. 

\item[3.] Weigh the dry sediment sample with the scale to obtain the total mass $M$. Consider an amount of sediment that allows for visual identification of the bed variation ($5-10\,\mathrm{cm}$). 

\item[4.] Put the sediment into the container and pour water in it. Measure the water height $H$. 

\item[5.] Prepare to take measurements of the bed height in time. Mix (shake) the sample and leave the container at rest and start measuring $z_{\text{bed}}(t_i)$ at different times $t_i$ (Figure~\ref{fig:experiment}). At the beginning, the sampling rate should be higher, since the sedimentation rate is higher due to higher suspended sediment concentration. The adequate sampling rate depends much on the sediment size. 

\item[6.] Measure the bed height $H_D$ when all the sediment is deposited at the bottom. Estimate the volume fraction of sediment in water $\phi_s$ from Equation~\ref{eq:HD}. 

\item[7.] Repeat items [5-6] at least three times. This allows to estimate the uncertainty of the measurements. 

\item[8.] Plot the observed values of the bed height as a function of time $\{t_i,z_{\text{bed}}(t_i)\}$. Estimate the best fit value for the effective settling velocity $\omega_s$ and, by using that value, plot the exact and approximate solutions. The Python code 'settling.py' can be used for that (Supplementary Material; see Appendix A).

\end{enumerate}

To evaluate the goodness of the approach, two experiments with two well-sorted sediment samples were carried out: (a) Experiment 1 with a sieved sample with sediment diameters within the range $[0.250,\,0.500]\,\mathrm{mm}$ (medium sand); (b) Experiment 2 with a sieved sample with sediment diameters in the interval $[0.125,\,0.250]\,\mathrm{mm}$ (fine sand). 

Figure~\ref{fig:experiment} (panels (a) and (b)) shows the measurements of $z_{\text{bed}}(t)$ corresponding to Experiments 1 and 2, respectively. For both experiments, the observations show higher sedimentation rates during the first stages of the process due to the higher suspended sediment concentration. As expected, the bed height reaches its maximum value $H_D$ sooner in the case of medium sand (panel (a)), i.e., suspended sediment concentration of medium sand exhibits a shorter $e$-folding time than fine sand. Sedimentation rates slow down when suspended concentration decreases as the sediment settles. 

The agreement between observations and the model solution is good, in particular for medium sand (panel (a)). The values of the effective settling velocity $\omega_s$ fitted are $\omega_s=\left(6.9\pm 0.3\right)\,\mathrm{cm\,s^{-1}}$ and $\omega_s=\left(3.3\pm 0.3\right)\,\mathrm{cm\,s^{-1}}$ for Experiment 1 and 2, respectively. These values are among the range given by the Udden-Wentworth grain size chart \citep{Wentworth1922} for medium and fine sand, i.e., $\sim 3.2-7.8\,\mathrm{cm\,s^{-1}}$ and $\sim 1.1-3.2\,\mathrm{cm\,s^{-1}}$, respectively. Table~\ref{tab:tabla} summarizes the results. Noticeable deviations between observations and model occur when most of the sediment is deposited ($\approx 80\%$): higher measured bed heights than the ones predicted by the model solutions are observed. This is more evident in the case of Experiment 2 (fine sand). This is likely due to the reduction of the hindered settling effects at lower suspended sediment concentrations. This effect, which has not been taken into account explicitly in the derivations, is more noticeable for fine sediment samples (panel (b)).

\begin{figure}
\centering{{\includegraphics[width=8cm]{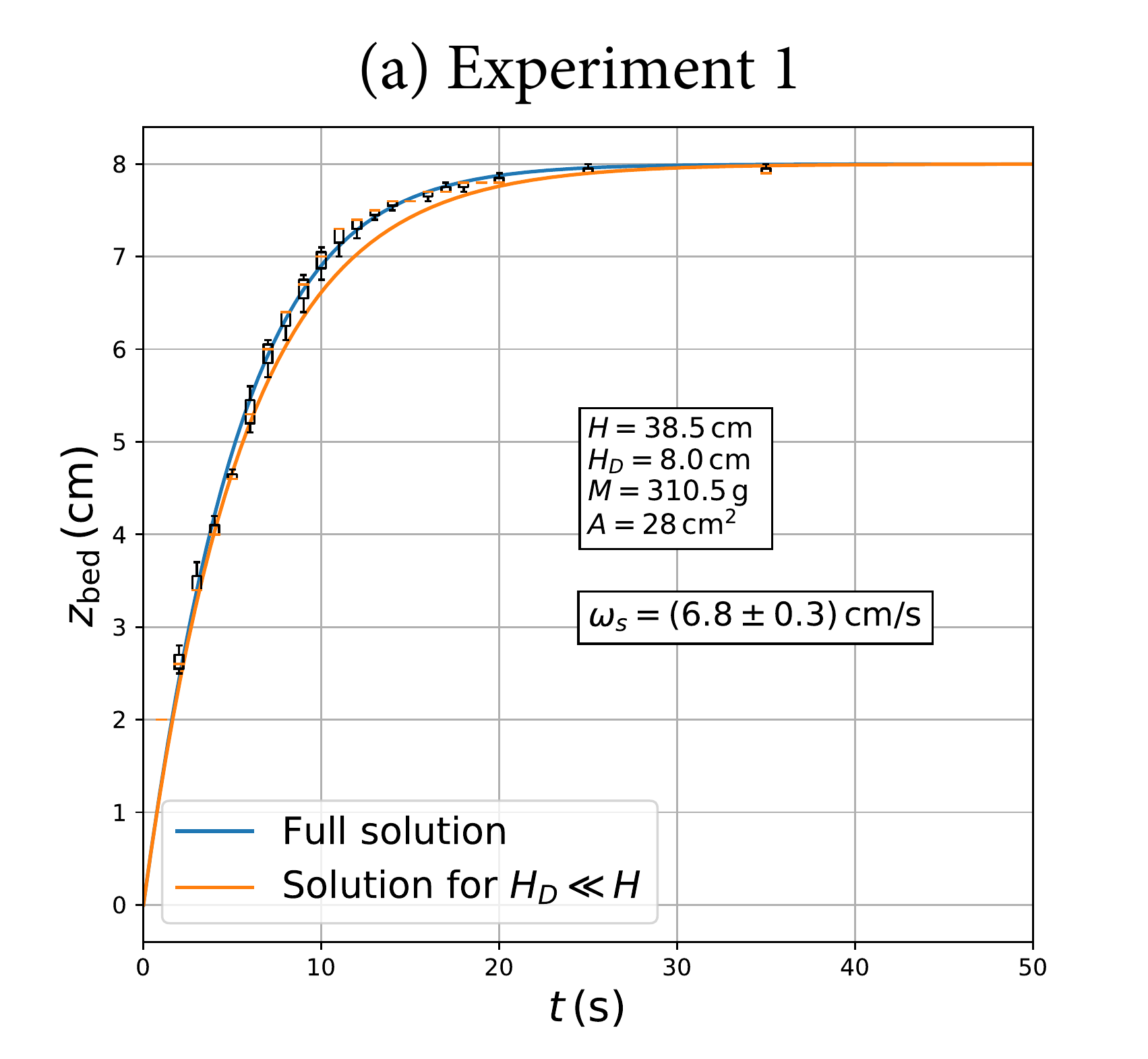}}}
\centering{{\includegraphics[width=8cm]{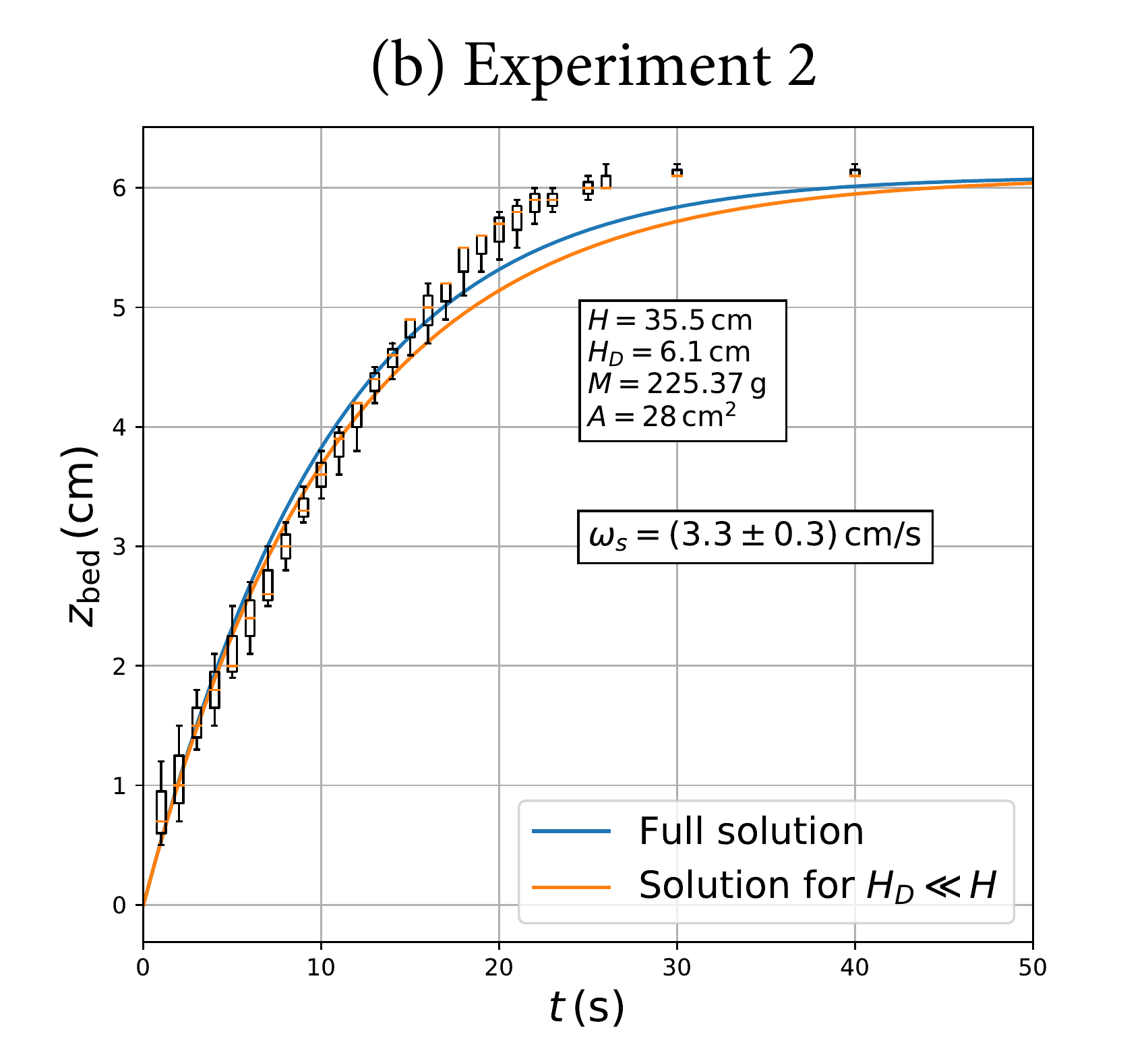}}}
\caption{\footnotesize{Time evolution of the bed height for Experiment 1 with medium sand (left panel (a)) and Experiment 2 with fine sand (right panel (b)). The values of the parameters $H$, $H_D$, $M$, and $A$ for each experiment are indicated in the boxes. Observations are depicted with dots and error bars. Blue solid curves are best fit full solutions of Equations~\ref{eq:eqverticalintegrada2} and \ref{eq:dzdt} (both panels). The best-fit settling velocities are $\omega_s=\left(6.8\pm 0.3\right)\,\mathrm{cm/s}$ and $\omega_s=\left(3.3\pm 0.3\right)\,\mathrm{cm/s}$ for Experiment 1 and 2, respectively. Orange solid curves are approximate solutions assuming $z_{\text{bed}} \ll H$} for the same $\omega_s$ values. } \label{fig:fit}
\end{figure}


\begin{table}[!htbp]
\label{tab:tabla}
\centering
\caption{Results of the Experiments. The effective settling velocities $\omega_s$ are the best fit values obtained using a standard least squares method.}
\begin{tabular}{l *2c}
\hline
{} & Experiment 1 & Experiment 2\\
\hline
$A\,\left(\mathrm{cm^2}\right)$ & $27.78\pm 0.01$   & $27.78\pm 0.01$ \\
$M\,\left(\mathrm{g}\right)$   &  $310.5\pm 0.1$ & $225.4\pm 0.1$ \\
$H\,\left(\mathrm{cm}\right)$   & $38.5\pm 0.1$ & $35.5\pm 0.1$ \\
$H_D\,\left(\mathrm{cm}\right)$   &  $8.0\pm 0.1$ & $6.1\pm 0.1$  \\ \hline
$\omega_s\,\left(\mathrm{cm\,s^{-1}}\right)$   &  $6.8\pm 0.3$  &  $3.3\pm 0.3$ \\
$\phi_s$   &  $0.523\pm 0.007$  &  $0.498\pm 0.009$ \\
\hline
\end{tabular}
\end{table}

\newpage

\section{Final Remarks}
This work allows to introduce the students into the modeling of sediment transport processes using simple physical and mathematical methodologies. It is easy to extend the experimental procedure for its study using salt water, which would increase the ambient fluid concentration and would favor the formation of flocs. Thus, a comparison between fresh and salt water settling velocities could be made. The overall good agreement between the idealized model and the experimental results encourages the students to develop simple models, analyze reasonable approximations and focus on the key physical process that describe the observations. The model is flexible enough to include, for instance, erosion rates or hindered settling effects and play with the model seeking stationary solutions. This can open up discussions about the role of physical forcings, such as wind, tides, on the erosion-deposition processes in real-world environments (e.g., rivers, estuaries, and coastal areas). This activity provides the students with competences for the development of physical and mathematical models based on conservation equations, which are key in many scientific fields in physics, engineering, and earth sciences.

\bibliography{bibliografia}

\appendix*
\section{Description of the Supplementary Material}\label{app:sup}

The supplementary material is comprised by data files and a Python 3.7 code that solves the ODE system defined by Equations~\ref{eq:eqverticalintegrada2} and \ref{eq:dzdt}. Python code and data files can be downloaded from \url{https://nasgdfa.ugr.es:5001/sharing/hadagOqsa}\\

Python 3.7 is an interpreted scripting language. Python is used for numerical programming and plotting the results (\url{https://www.python.org/}). Packages NumPy (\url{http://www.numpy.org/}) and SciPy (\url{https://www.scipy.org/}) are used for integrating the differential equations and optimizing parameters by least squares fits to data. In particular, the Scypy function \textit{odeint()} is used to solve systems of ODEs. This function is based on the LSODA code from the FORTRAN library odepack. Package matplotlib is used for plotting (\url{https://matplotlib.org/}). 

Two sample data files are provided. Data files '\textit{z\_d\_t\_exp1.dat}' and '\textit{z\_d\_t\_exp2.dat}' contain observations of bed height in time for sediment samples 1 (medium sand) and 2 (fine sand), respectively. The Python code '\textit{settling.py}' implements the standard least squares method to determine settling velocity $\omega_s$, and the functions to solve the ODE system defined by Equations~\ref{eq:eqverticalintegrada2} and \ref{eq:dzdt}. The code also plots observations and the numerical and approximate solutions. More details are given as in-line comments within the code.

\begin{acknowledgments}

The authors acknowledges support from Programa Estatal de I+D+i RETOS (Ref. CTM2017-89531-R).

\end{acknowledgments}

\end{document}